\documentclass[twoside]{article}
\usepackage{fleqn,espcrc2}

\usepackage{epsfig}

\newcommand{\beq}{\begin{equation}}
\newcommand{\eeq}{\end{equation}}
\newcommand{\bea}{\begin{eqnarray}}
\newcommand{\eea}{\end{eqnarray}}

\newcommand{\lsim}{\raisebox{-0.04cm}{$\:\stackrel{<}{{\scriptstyle
 \sim}}\: $} }

\title{Real photon structure at an $e^+e^-$ linear collider}

\author{Andreas Vogt%
\address{Instituut-Lorentz, University of Leiden, P.O. Box 9506,
 2300 RA Leiden, The Netherlands}%
\thanks{Work supported by the EC network `QCD and Particle Structure'
 under contract No.~FMRX--CT98--0194.}}
       
\begin{document}

\begin{abstract}
Previous studies of the kinematic coverage for measuring the photon 
structure function $F_2^{\,\gamma}$ at a future 500~GeV $e^+e^-$ linear 
collider \cite{LC1,LC2} are updated using current estimates of  
luminosities and important detector parameters. The perturbative 
expansion for the evolution of $F_2^{\,\gamma}$ is briefly recalled in 
view of a recent claim \cite{Ch98} that all existing next-to-leading 
order analyses of the photon structure are incorrect. A simple 
illustration is given of the different sensitivities of hadronic and 
photonic structure functions on the strong coupling constant 
$\alpha_s$.
\vspace{1pc}
\end{abstract}

\maketitle

\section{KINEMATICS}

\begin{figure}[thb]
\label{avf1}
\vspace*{-17mm}
\centerline{\epsfig{file=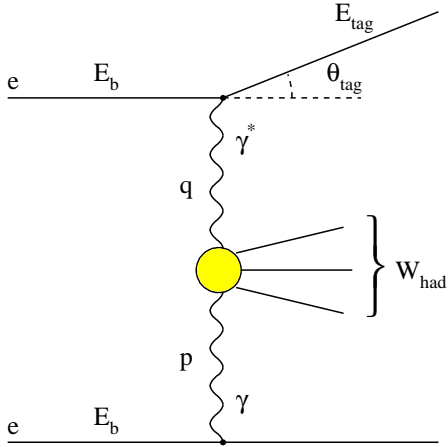,width=6.3cm,angle=0}}
\vspace*{-10mm}
\caption{The kinematics of inclusive  electromagnetic electron 
 scattering off (quasi-)real photons in $e^+e^-$ or $e^-e^-$ 
 collisions.}
\vspace*{3mm}
\end{figure}

The basic kinematics of electron-photon ($e\gamma$) deep-inelastic 
scattering (DIS) at an $e^+ e^-$ or $e^- e^-$ collider is recalled in 
Fig.~\ref{avf1}. Here and in what follows we consider only the 
electromagnetic one-photon-exchange process, i.e., it is assumed that 
QED radiative corrections and contributions due to the exchange of weak 
gauge bosons have been subtracted. For a recent study of neutral- and
charged-current processes see ref.~\cite{GDR}.

The unpolarised $e\gamma$ DIS cross section is, at lowest order in 
the electromagnetic coupling $\alpha$,
\bea
\label{ave1}
 \lefteqn{
  \frac{d \sigma (e \gamma \rightarrow eX)}{dE_{\rm tag}\, d\cos 
  \theta_{\rm tag}}\: = \:\frac{4\pi \alpha^2 E_{\rm tag}}{Q^4 y} 
  \:\cdot } \\ & & \cdot
  \Big[ \big\{ 1 + (1-y)^2 \big\} F_2^{\, \gamma}(x,Q^2) 
  - y^2 F_L^{\, \gamma}(x,Q^2) \Big] \: . \nonumber
\eea
Here $F_{2,L}^{\, \gamma}(x,Q^2)$ denote the structure functions of the
real photon. The virtuality of the probing photon and the invariant 
mass of the hadronic final state are given by
\bea
\label{ave2}
  Q^2 \:\: & \equiv\! & -q^2 \: = \: 2 E_b E_{\rm tag} 
  (1-\cos\theta_{\rm tag}) \: ,  \nonumber \\
  W^2_{\rm had} \!\! & =\! & (q+p)^2 \: .
\eea
The scaling variables $x$ and $y$ read
\bea
\label{ave3}
  x &\! =\! & \frac{Q^2}{Q^2+W^2_{\rm had}} \: ,  \nonumber \\
  y &\! =\! & 1- \frac{E_{\rm tag}}{E_b}
      \cos^2 \left( \frac{\theta_{\rm tag}}{2} \right) \: .
\eea
The measured cross section is obtained by convoluting Eq.~(\ref
{ave1}) with the flux $f_{\gamma ,e}(z\! =\! E_{\gamma}/E_{\rm beam})$ 
of the incoming (quasi-)real photons.

If the photon momentum $p$ is known, then not only $Q^2$ and $y$, but
also $W_{\rm had}$ and $x$ in Eqs.~(\ref{ave2}) and (\ref{ave3}) are
fixed by energy and angle of the `tagged' outgoing electron, as in
usual electron-proton DIS. If $p$ is unknown, the determination of $x$ 
has to proceed via calorimetric measurements of the hadronic final 
state. Beam-pipe losses then render it very difficult, if not
impossible, to obtain high-precision data on $F_2^{\,\gamma}(x,Q^2)$. 

\section{PHOTON SPECTRA}

A ubiquitous source of quasi-real photons at an electron collider is 
the soft bremsstrahlung (`Weiz\-s\"acker-Williams') spectrum emitted by 
almost undeflected electrons \cite{BGMS}. This spectrum has been the 
photon source for the $F_2^{\,\gamma}$ measurements at LEP and previous 
$e^+e^-$ colliders \cite{LEP}, where the corresponding electrons are
undetected (`anti-tagged'). In order to determine the momenta of the 
quasi-real photons, however, as required for precision measurements of 
$F_2^{\,\gamma}$, these forward electrons need to be tagged as well. 
Due to the high beam energies $E_b$ at a linear collider this should be 
done at angles $\theta_f \lsim 2$ mrad, restricting the high-virtuality 
tail of the WW spectrum which reaches up to $ P^2 \equiv -p^2 \simeq 
(1-z) \, E_b^{\, 2} \, \theta_f^ {\, 2}$. It seems not too likely that 
this can be achieved with high accuracy and efficiency. 

For an optimistic estimate of possible event numbers in this scenario 
we will assume a 10\% efficiency for $0.2\, E_b \leq x \leq 0.8\, E_b$.
Thus we use
\bea
\label{ave4}
 \lefteqn{
 f_{\gamma /e}^{\rm WW}(z) = K \,\theta(z-0.2) \,\theta (0.8-z) 
 } \\ & & \!\!\!
  \frac{\alpha}{\pi z}
  \Big[ \big( 1 + (1-z)^2 \big ) \ln \frac{E_b \theta_f (1-z)}{m_e z} 
      - (1-z) \Big] \:  \nonumber
\eea
with $K = 0.1$ and $\theta_f = 2$ mrad.

A very appealing new possibility envisaged for the linear collider is
to operate the machine as an $e\gamma $ collider as well. This can be 
realized by converting one of the electron beams to a real-photon beam 
by backscattering of laser photons \cite{BL1}. Subsequently the 
resulting broad spectrum can be transformed into a rather monochromatic 
photon beam, 
\beq 
\label{ave5}
   E_{\gamma} \simeq 0.8\, E_{b}\: $ with $\:\:\Delta E_{\gamma} 
\approx 0.1 \, E_{\gamma} \: ,
\eeq 
under suitable machine conditions, see ref.~\cite{BL2}.

\begin{figure}[ht]
\label{avf2}
\centerline{\epsfig{file=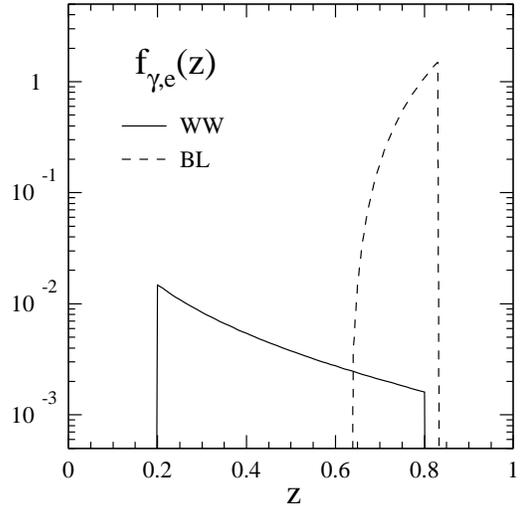,width=6.8cm,angle=0}}
\vspace*{-8mm}
\caption{The effective flux functions for bremsstrahlung (WW) and 
 backscattered laser (BL) photons given in Eq.~(\ref{ave4}) and
 Eq.~(\ref{ave6}), respectively.}
\end{figure}

For our illustrations below we will employ a simple model spectrum
\cite{LC2} incorporating Eq.~(\ref{ave5}) and the rough shape of the 
high-$z$ peak: 
\bea
\label{ave6}
 \lefteqn{
 f_{\gamma /e}^{\rm BL}(z) = K' \theta(z\! -\! 0.63) 
 \,\theta (0.83\! -\! z)\, 375 \, (z\! -\! 0.63)^2 } \nonumber \\
 & & 
\eea
with $ K' = 0.1$. The latter suppression factor leads to a conservative 
estimate of the attainable $e\gamma$ luminosity. The flux functions 
(\ref{ave4}) and (\ref{ave6}) are compared in Fig.~\ref{avf2}.

\section{EVENT NUMBERS, \boldmath{$F_2^{\,\gamma}$} COVERAGE}

The kinematic coverage for measurements of $F_2^{\,\gamma}(x,Q^2)$
strongly depends on the minimal angle $\theta_{\rm tag,min}$ down to 
which the scattered electron can be detected. In fact, an electron 
tagger inside the radiation shielding masks of the main detector at 
about 10 degrees is indispensable for accessing the region $Q^2 < 1000$ 
GeV$^2$ at a 500~GeV machine. Presently $\theta_{\rm tag,min} = 25$ 
mrad is considered feasible, a value even below the 40 mrad demanded in 
ref.~\cite{LC2}.
Also important is the background suppression cut $E_{\rm tag,min}$ on
the energy of the scattered electron. For the present study we replace 
the previous choice $0.5\, E_b$ by the weaker requirement 
$E_{\rm tag,min} = 50$ GeV. The accessible $y$-range is thus enlarged 
from $y \leq 0.5$ to $y \leq 0.8$. 

\begin{figure}[t]
\label{avf3}
\vspace*{-1mm}
\centerline{\epsfig{file=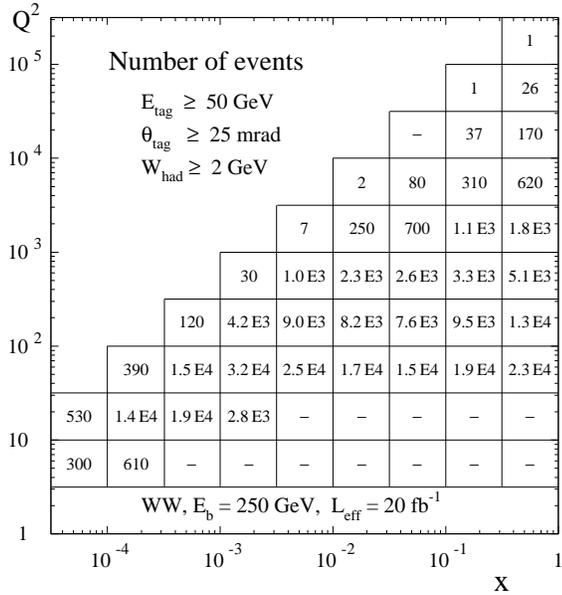,width=7.5cm,angle=0}}
\vspace*{-8mm}
\caption{Expected event numbers of $e\gamma $ DIS for bremsstrahlung 
 photons at a 500 GeV linear collider. Forward-electron tagging is 
 assumed to be possible with 10\% efficiency for $\theta_f \leq 2$ mrad 
 and electron energies between 50 and 200 GeV.}
\vspace*{-2mm}
\end{figure}

The resulting event numbers are shown in Fig.~\ref{avf3} for the 
bremsstrahlung scenario (\ref{ave4}), and in Fig.~4 for the laser 
backscattering scenario~(\ref{ave6}). $F_{2.L}^{\,\gamma}$ in 
Eq.~(\ref{ave1}) are calculated using the leading-order GRV 
parametrisation~\cite{GRVg}. An integrated $e^+e^-$ luminosity of 200 
fb$^{-1}$, typical for the present {\sc Tesla} design, is assumed. In 
both cases this luminosity is effectively reduced by a factor of 10 by 
the $f_{\gamma,e}$ assumptions in Sect.~2. 
Under these conditions the $e\gamma$ collider is, as expected from 
Fig.~\ref{avf2}, vastly superior to the WW scenario. E.g., the 
difference in the high-$Q^2$ reach at large $x$ amounts to one order of 
magnitude.

\begin{figure}[t]
\label{avf4}
\vspace*{-1mm}
\centerline{\epsfig{file=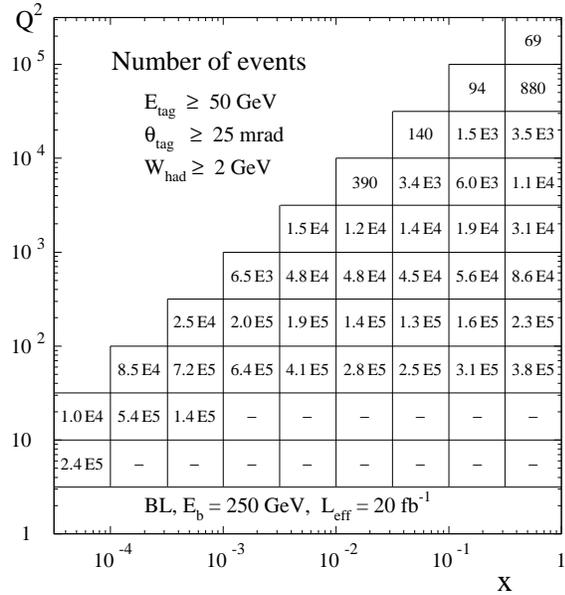,width=7.5cm,angle=0}}
\vspace*{-8mm}
\caption{Expected event numbers of electron-photon DIS for the 
 backscattered-laser $e\gamma$ mode of a 500 GeV linear collider. It is 
 assumed that 10\% of the $e^+ e^-$ luminosity can be reached in this 
 mode for a rather monochromatic photon beam.}
\end{figure}

The maximal accuracies for $F_2^{\,\gamma}$ determinations are 
illustrated in Fig.~5 for the preferred $e\gamma$ case. Here we have 
assumed that the systematic error is equal to the statistical one 
inferred from the event numbers, but amounts to at least 3\%. For the 
present parameters, precision measurements are possible at large $x$ 
for $40 \mbox{ GeV}^2 \lsim Q^2 \lsim 10^4 \mbox{ GeV}^2 $. The region 
$Q^2 < 30 \mbox{ GeV}^2$ can only be reached by a (preferably 
asymmetric) lower energy run. 
However, such $Q^2$-values can be accessed at very small $x$. Only in
this region, $x \approx 10^{-4}$, the cross section (\ref{ave1}) 
receives noticeable contributions from the longitudinal structure 
function $F_L^{\,\gamma}$. 

Charm production contributes about 30--40\% to the cross section over 
almost the entire kinematic region of Fig~4. A decent measurement of 
$F_2^{\,\rm charm}$ should thus be possible at large $x$, where the 
final-state particles are not too forward.

\begin{figure*}[t]
\label{avf5}
\centerline{\epsfig{file=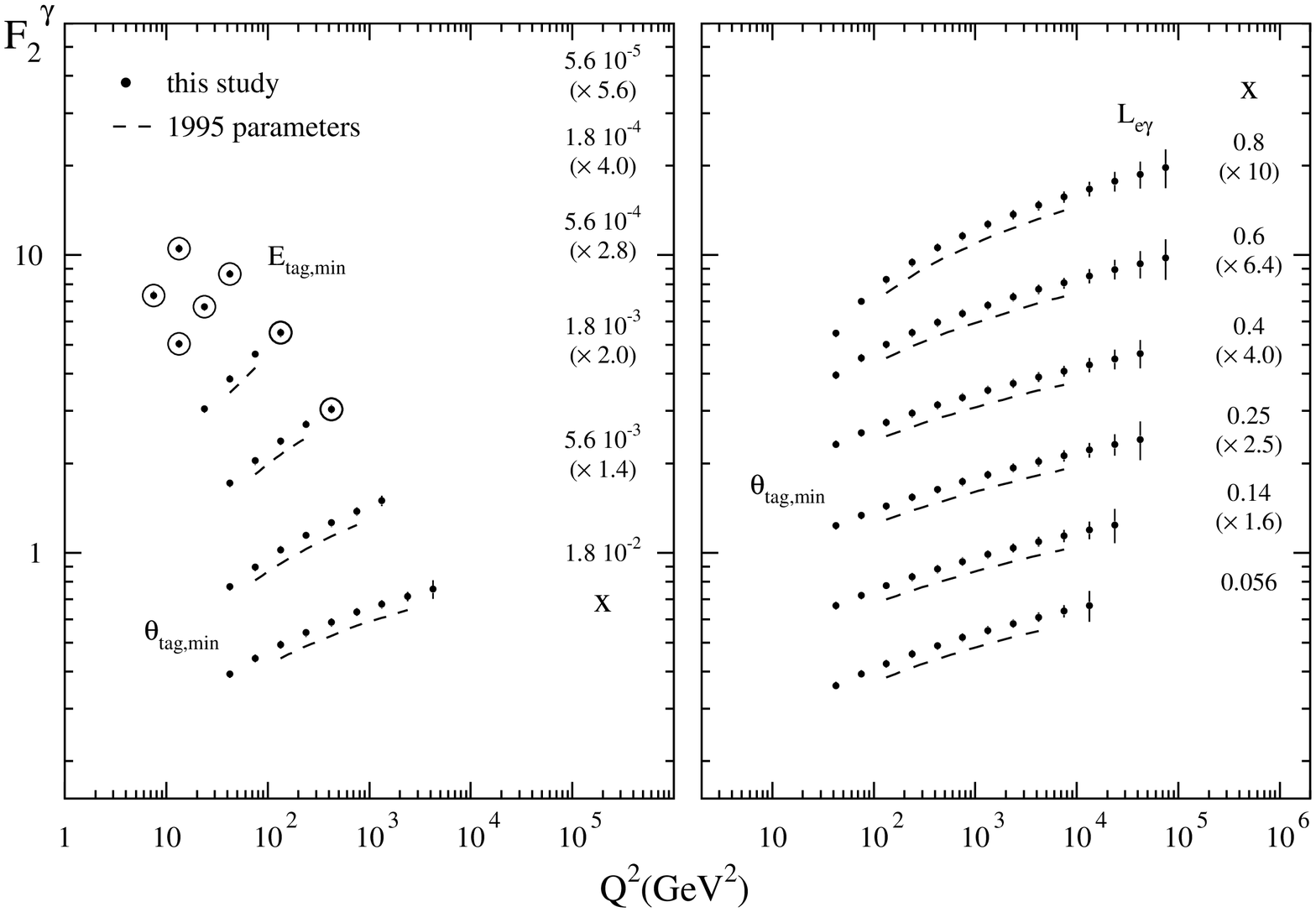,width=15.0cm,angle=0}}
\vspace*{-8mm}
\caption{The possible kinematic coverage and maximal accuracy of the
 measurement of $F_2^{\,\gamma}$ for the backscattered-laser $e\gamma$ 
 mode at a 500 GeV linear collider. The open circles show the bins 
 with an expected $F_L^{\,\gamma}$-effect of 10\% or more. The sources 
 of the extended $x$ and $Q^2$ reach with respect to the 1995 
 study~\cite{LC2} are indicated. The values of $F_2^{\,\gamma}$ have
 been scaled by the factors given in the figure.} 
\end{figure*}

\section{EVOLUTION OF \boldmath{$F_2^{\,\gamma}$}}

We now turn to the theoretical description of $F_2^{\,\gamma}$ in 
perturbative QCD.  For notational simplicity we restrict ourselves to 
the flavour non-singlet part, and consider the contributions of 
effectively massless quarks only (see ref.~\cite{BVpr} for the singlet 
case). All functional dependences and obvious indices will be 
suppressed together with overall charge factors, e.g., we will write 
$F^\gamma$ instead of $F_{2,\rm NS}^{\,\gamma}(x,Q^2)/\alpha $. It is 
understood that all products of $x$-dependent quantities have to be 
read as convolutions in Bjorken-$x$ space.

In terms of a suitable combination $q^\gamma$ of quark densities the 
non-singlet structure function reads
\beq
\label{ave7}
  F^\gamma = C_q \, q^{\gamma} + C_\gamma
\eeq
at lowest order in the electromagnetic coupling $\alpha$. In this
approximation the quark density $q^\gamma$ is subject to the 
inhomogeneous evolution equation
\beq
\label{ave8}
  \dot{q}^\gamma \equiv \frac{dq^\gamma}{d \ln Q^2} 
  = k + P \, q^{\gamma} 
\eeq
for the choice $\mu_f^2 = Q^2$ of the factorization scale.

The corresponding photon and quark coefficient functions $C_q$ and 
$C_\gamma$ in Eq.~(\ref{ave7}) have the perturbative expansions
\bea
\label{ave9}
  C_q = 1 + \sum_{l=1} a_s^l C_q^{(l)} \: , \:\:\:
  C_\gamma = \sum_{l=1} a_s^{l-1} C_\gamma^{(l)}
\eea
in terms of the strong coupling $a_s \equiv \alpha_s / (4\pi)$. The 
scale dependence of $a_s$  is governed by
\beq
\label{ave10}
  \frac{da_s}{d \ln \mu_r^2} = \beta (a_s) = - \sum_{l=0} a_s^{l+2} 
  \beta_l
\eeq
where $\mu_r$ is the renormalization scale. Finally the non-singlet 
quark-quark and photon-quark splitting functions in Eq.~(\ref{ave8})
read
\beq
\label{ave11}
  P = \sum_{l=0} a_s^{l+1} P^{(l)} \: , \:\:\: 
  k = \sum_{l=0} a_s^l k^{(l)} \:\: .
\eeq

The expansion coefficients in Eqs.~(\ref{ave9}) and (\ref{ave11}) are 
$Q^2$-independent for $\mu_r^2 = \mu_f^2 = Q^2$. They do depend,
however, on the mass-factorization and renormalization schemes. While
the latter dependence is unavoidable in perturbative calculations, the 
former can be eliminated by expressing the observable $\dot{F}^\gamma 
\equiv dF^\gamma / d\ln Q^2$ in terms of $F^\gamma$ itself. By 
inserting Eq.~(\ref{ave7}) solved for $q_\gamma$ into
\beq 
\label{ave12}
  \dot{F}^\gamma = \dot{C}_q \cdot q^{\gamma} + C_q \cdot 
  \dot{q}^{\gamma} + \dot{C}_\gamma
\eeq
after using Eq.~(\ref{ave8}), one arrives at 
\bea
\label{ave13}
  \dot{F}^\gamma &\!\! = &\!\!  
  \big\{ - \dot{C}_q C_q^{-1} C_\gamma + C_q k - P C_\gamma + 
  \dot{C}_\gamma \big\} \nonumber \\
  & & \mbox{} + \big\{ \dot{C}_q C_q^{-1} +  P \big\} \,
  F^\gamma \:\: .
\eea
The combinations of splitting functions and coefficient functions 
enclosed by the brackets are scheme invariant. 

Insertion of the perturbative expansions (\ref{ave9}--\ref{ave11}) into 
Eq.~(\ref{ave13}) finally yields, for $\mu_r^2 = Q^2$,

\noindent
\bea
\label{ave14}
  \dot{F}^\gamma \!\! &\!\! = & \! a_s^0 \: k^{(0)} \nonumber \\
  & & \mbox{} \hspace*{-5mm} 
     + a_s^1 \big\{ k^{(1)} + C_q^{(1)} k^{(0)} - P^{(0)} 
     C_\gamma^{(1)} \big\} \nonumber \\
  & & \mbox{} \hspace*{-5mm} 
     + a_s^2 \big\{ k^{(2)} +  C_q^{(1)} k^{(1)} + C_q^{(2)} k^{(0)}
     - P^{(0)} C_\gamma^{(2)} \nonumber \\
  & & \mbox{}\quad - P^{(1)} C_\gamma^{(1)} - \beta_0 
     C_\gamma^{(1)} - \beta_0  C_q^{(1)} C_\gamma^{(1)} \big\} 
     \nonumber \\
  \!\! +\!\!\! & \!\! F^\gamma \! & \!\!\!\!\Big[ \, a_s^1\: P^{(0)} \\
  & & \mbox{} \hspace*{-5mm} 
     + a_s^2 \big\{ P^{(1)} - \beta_0 C_q^{(1)} \big\} \nonumber \\
  & & \mbox{} \hspace*{-5mm}
     + a_s^3 \big\{ P^{(2)} - \beta_0 [ 2 C_q^{(2)} - C_q^{(1) 2} ] 
     - \beta_1 C_q^{(1)} \big\} \Big] \nonumber \\
  & & \mbox{} \hspace*{-5mm} + O(a_s^3)\: + \: F^\gamma \cdot O(a_s^4) 
      \nonumber \:\: .
\eea
The generalization to $\mu_r^2 \neq Q^2$ can be obtained by expressing 
$a_s \equiv a_s(Q^2)$ as a power series in $a_s(\mu_r^2)$ via 
Eq.~(\ref{ave10}), inserting this expression into Eq.~(\ref{ave14}), 
and re-expanding the result in terms of $a_s(\mu_r^2)$ up to the same
order as before for $a_s(Q^2)$.
The combinations enclosed in the curved brackets in Eq.~(\ref{ave14})
are, as the leading terms $k^{(0)}$ and $P^{(0)}$, factorization scheme 
(and scale) invariant since the relation (\ref{ave14}) between
observables holds irrespective of the values of $a_s$ and $F^\gamma$.

Note that $C_\gamma^{(1)}$, despite being an $a_s^0$-coefficient in 
Eq.~(\ref{ave9}), enters $\dot{F}_2^\gamma$ on the same level as the 
hadronic next-to-leading order (NLO) quantity $C_q^{(1)}$. Hence 
$C_\gamma^{(1)}$ is to be considered as a NLO contribution as well. 
Notice also that $k^{(1)}$, like its hadronic NLO  counterpart 
$P^{(1)}$, is changed by a re-definition of $C_q^{(1)}$, e.g., by 
switching from the $\overline{\mbox{MS}}$ or DIS$_\gamma$ schemes 
\cite{evol} to the DIS scheme. Correspondingly $C_\gamma^{(2)}$ and the 
presently uncalculated quantity $k^{(2)}$ contribute in 
next-to-next-to-leading order only. 
Looking back to Eqs.~(\ref{ave8}) and Eqs.~(\ref{ave14}) these findings
imply that, for the purpose of power-counting in $\alpha_s$, the quark
densities $q^\gamma$ and the structure function $F_2^{\,\gamma}$ have 
to be counted as $1/\alpha_s$. 
 
The correct counting just described is implemented in the NLO 
photon-structure parametrisations of refs.~\cite{GRVg,prms} and 
countless applications to specific processes. It is, however, in 
direct contradiction to ref.~\cite{Ch98}, where $C_\gamma^{(1)}$ is 
considered as a leading-order term and $C_\gamma^{(2)}$ and $k^{(2)}$ 
are claimed to be NLO quantities.

\section{SENSITIVITIES ON {\large\boldmath{$\alpha_s$}}}

The dependence of the large-$x$ evolution of $F_2^{\,\gamma}$ on the
strong coupling constant $\alpha_s$ is quite different from that of
hadronic structure functions. The latter case is recalled in Fig.~6
where the NLO pion parametrisation of ref.~\cite{GRVm}, after 
transformation to the DIS scheme at $Q^2 = 4$ GeV$^2$, is evolved for 
four flavours. The $\alpha_s$-sensitive quantities are the logarithmic 
$Q^2$-slopes of $F_2$ at large $x$; an overall normalization error is 
irrelevant.

The corresponding results for the photon case are illustrated in 
Fig.~7 using the parametrisation of ref.~\cite{GRVg}. Here the 
$\alpha_s$-sensitive quantities are the absolute values of 
$F_2^{\,\gamma}$ at very large-$x$ and high $Q^2$, cf.\ 
ref.~\cite{Fr87}. At $x = 0.8$, e.g., a change of $\alpha_s(M_Z)$ by 
5\% leads to an effect of about 3\% on $F_2^{\,\gamma}$.  

\begin{figure}[bht]
\label{avf6}
\vspace*{-1mm}
\centerline{\epsfig{file=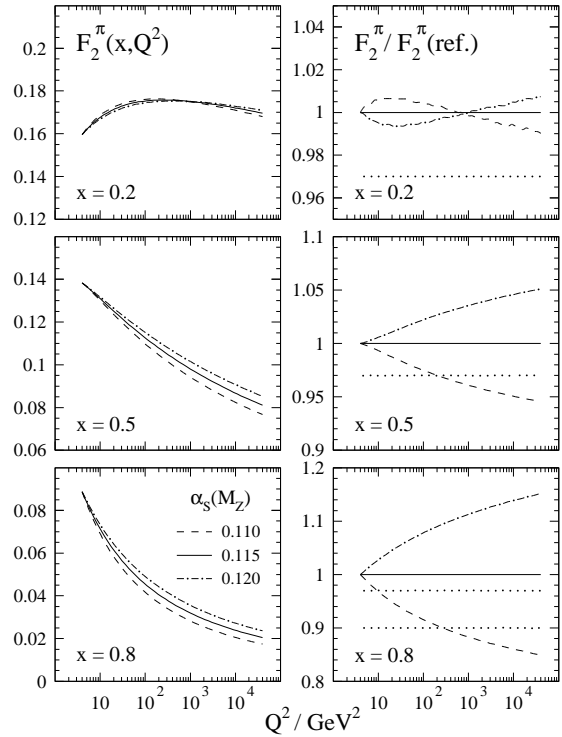,width=7.3cm,angle=0}}
\vspace*{-9mm}
\caption{The $\alpha_s$-dependence of the large-$x$ NLO evolution of 
 the neutral-pion structure function $F_2^{\,\pi}$ for a fixed input at
 $Q^2 = 4$ GeV$^2$. The dotted lines show the effect of input 
 normalization offsets of $-3\%$ and $-10\%$ on the $\alpha_s(M_Z)\!
 =\! 0.115$~curves.}
\end{figure}

\begin{figure}[thb]
\label{avf7}
\centerline{\epsfig{file=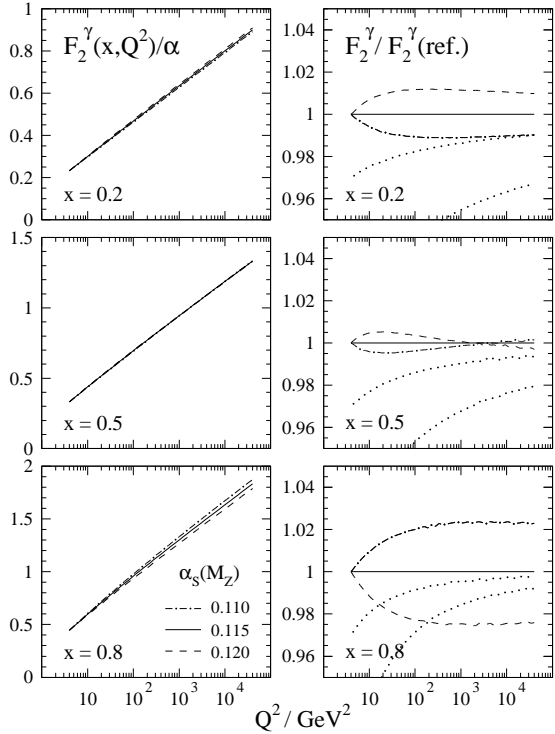,width=7.3cm,angle=0}}
\vspace*{-9mm}
\caption{As Fig.~6, but for $F_2^{\,\gamma}$. Note the 
 shrinking high-$Q^2$ impact of low-scale normalization changes at 
 large $x$, and the almost identical shapes of the upper dotted 
 ($\alpha_s(M_Z)\! =\! 0.115$) and the dash-dotted ($\alpha_s\! =\! 
 0.110$) curves at $x = 0.8$.}
\vspace*{-4mm}
\end{figure}

\section{CONCLUSIONS}

A high-energy $e\gamma$ collider, realised via Compton backscattering
of laser photons at a future $e^+e^-$ collider, would be a unique 
source of information on the photon structure. At a 500 GeV machine the 
structure function $F_2^{\,\gamma}$ can be accurately measured at large 
$x$ for $ 40 \mbox{ GeV}^2 \lsim Q^2 \lsim 10^4 \mbox{ GeV}^2$. Small 
values of $x$ down to below $10^{-4}$ can be reached for $Q^2 \approx 
10 $ GeV$^2$, thus complementing the $F_2^{\, p}$ results of HERA in 
this regime.

At large values of $x$ and $Q^2$, $x\! >\! 0.7$ and $Q^2>10^3$ GeV$^3$,
the absolute values of $F_2^{\,\gamma}$ become robust predictions of 
perturbative QCD. A competitive extraction of $\alpha_s$ from these
values would require very high accuracies on both the experimental and 
the theoretical sides.

\section*{ACKNOWLEDGMENT}

It is a pleasure to thank A. de Roeck for discussions on detector
parameters and final-state cuts for electron-photon DIS at the linear
collider.


\vspace{2mm}
\begin{thebibliography}{99}
\bibitem{LC1}  D.J. Miller et al., Proceedings of the Workshop on 
               Linear $e^+ e^-$ Colliders, 
               Waikaloa, Hawaii, April 1993, eds.\ F.A. Harris et al. 
               (World Scientific 1993), p.\ 577.
\bibitem{LC2}  D.J. Miller and A. Vogt, Proceedings of the Workshop on 
               $e^{+}e^{-}$ Collisions at TeV Energies (Annecy, Gran 
               Sasso, DESY 1995), ed.\ P. Zerwas (DESY 1996), p.\ 473;
               \\
               E. Accomando et al., Phys.\ Rep.\ 299 (1998)~1.
\bibitem{Ch98} J. Chyla, {\tt hep-ph/9811455}.
\bibitem{GDR}  A. Gehrmann-De Ridder, {\tt hep-ph/9906547} (these
               proceedings).
\bibitem{BGMS} V. Budnev et al., Phys.\ Rep.\ 15C (1975) 181.
\bibitem{LEP}  R. Nisius, {\tt hep-ex/9907012} (these proceedings), and
               references therein.
\bibitem{BL1}  I.F. Ginzburg et al., Nucl.\ Inst.\ Meth.\ A205 (1983) 
               47, A219 (1984) 5; \\
               V.I. Telnov, Nucl.\ Inst.\ Meth.\ A94 (1990) 72,
               {\tt hep-ex/9908005} (these proceedings). 
\bibitem{BL2}  R. Brinkmann et al., Nucl.\ Instrum.\ Meth.\ A406 (1998) 
               13. 
\bibitem{GRVg} M. Gl\"uck, E. Reya and A. Vogt, Phys.\ Rev.\ D46 (1992) 
               1973.
\bibitem{BVpr} J. B\"umlein and A. Vogt, Phys.\ Rev.\ D58 (1998) 014020.
\bibitem{evol} W.A. Bardeen and A.J. Buras, Phys.\ Rev.\ D20 (1979) 166,
               E: D21 (190) 2041; \\
               M. Fontannaz and E. Pilon, Phys.\ Rev.\ D45 (1992) 382,
               E: D46 (1992) 484; \\
               M. Gl\"uck, E. Reya and A. Vogt, Phys.\ Rev.\ D45 (1992)
               3986.
\bibitem{prms} L.E Gordon and J.K Storrow, Z. Phys.\ C56 (1992) 307, 
               Nucl.\ Phys.\ B489 (1997) 405; \\ 
               P. Aurenche, M. Fontannaz and J.P. Guillet, Z. Phys.\ 
               C64 (1994) 621; \\  
               M. Gl\"uck, E. Reya and I. Schienbein, Phys.\ Rev.\ 
               D60 (1999) 054019.
\bibitem{GRVm} M. Gl\"uck, E. Reya and A. Vogt, Z. Phys.\ C53 (1992) 
               652.
\bibitem{Fr87} W. R. Frazer, Phys.\ Lett.\ B194 (1987) 287.
\end{thebibliography}
\end{document}